\documentclass{article}
\usepackage{spconf,amsmath,graphicx,booktabs}
\usepackage{amsfonts}
\usepackage{multirow}
\usepackage{soul}

\title{INVESTIGATING PERSONALIZATION METHODS IN TEXT TO MUSIC GENERATION}

\name{%
\begin{tabular}{c}
Manos Plitsis $^1$ $^2$ $^*$, \qquad 
Theodoros Kouzelis $^1$ $^*$\thanks{$^{*}$ M. Plitsis and T. Kouzelis contributed equally} \\
Georgios Paraskevopoulos $^1$\qquad 
Vassilis Katsouros $^1$ \qquad 
Yannis Panagakis $^2$ $^3$
\end{tabular}}
\address{$^1$ Institute for Language and Speech Processing, Athena Research Center, Greece\\
  $^2$ Department of Informatics and Telecommunications, University of Athens, Greece \\
  $^3$ Archimedes Unit,  Athena Research Center, Greece
  }

\begin{document}
%
\maketitle
\begin{abstract}


In this work, we investigate the personalization of text-to-music diffusion models in a few-shot setting.
Motivated by recent advances in the computer vision domain, we are the first to explore the combination of pre-trained text-to-audio diffusers with 
two established personalization methods. We experiment with the effect of audio-specific data augmentation on the overall system performance
and assess different training strategies. For evaluation, we construct a novel dataset with prompts and music clips. 
We consider both embedding-based and music-specific metrics for quantitative evaluation, as well as a user study for 
qualitative evaluation. Our analysis shows that similarity metrics are in accordance with user preferences and that current
personalization approaches tend to learn rhythmic music constructs more easily than melody.
The code, dataset, and example material
of this study are open to the research community.

\end{abstract}
\section{Introduction}
\label{sec:intro}

Creating customized music and sound effects to meet individualized specifications can have significant impact across diverse application domains, including music production, augmented and virtual reality, and game development applications.
In recent years, there has been a growing number of text-to-music generative models \cite{liu2023audioldm,copet2023simple}. 
These models are versatile, capable of generating a diverse range of audio, including music, based on a textual prompt.


Guiding such models to a desirable output sound is not straightforward, requiring considerable prompt engineering \cite{gu2023systematic}.
This means that there is no way to finely control the generation process to consistently produce sounds based on a specific example.
This challenge arises either because the model cannot produce any instance of a class of sounds (e.g. an obscure ethnic instrument) or because the desired sound is a specific instance of a known class (e.g. producing a user's guitar playing style) that cannot be yielded even with the most detailed textual description.
 For instance, can one generate a rock song using their personal guitar playing style or a specific ethnic instrument?

\begin{figure}[t]
  \centering

\includegraphics[scale=0.3]{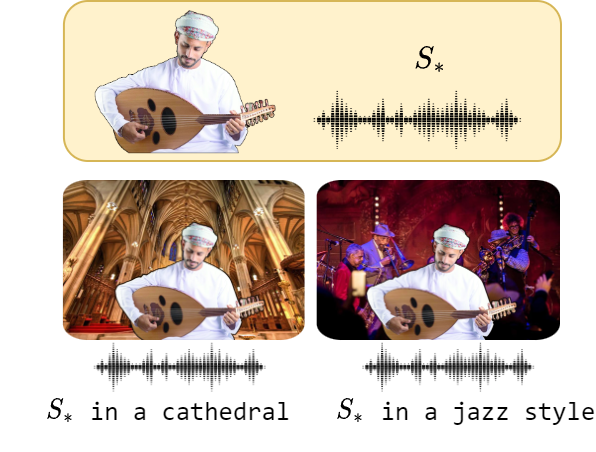}
  \caption{An overview of text-to-music personalization. With just a few audio clips, we implant a novel musical concept into a pre-trained text-to-audio model, enabling its manipulation with textual prompts.}
  \label{fig:overviw}
\vspace{-10 pt}
\end{figure}

In the image domain, this problem is addressed by \textit{personalization} methods that expand the language-vision dictionary of the model so that it binds new words with user-specific concepts. This enables the generation of a user-specific concept in different contexts and stylistic variations while maintaining its distinct characteristics \cite{ruiz2023dreambooth}.
Given a few examples, e.g. ${\sim} 3-5$ images of a dog in different backgrounds and views, the objective of personalization methods is to inject it into the model such that it can be synthesized with a unique identifier (e.g. a pseudoword).

 Recently, several approaches
based on pre-trained text-to-image diffusion models have been proposed \cite{gal2023an, ruiz2023dreambooth,kumari2022customdiffusion}.
 Textual Inversion \cite{gal2023an} adds a new word embedding for the novel concept and associates it with a pseudoword $V_*$. The embedding is trained with prompts of the form "a photo a $V_*$" via the standard denoising objective \cite{ldm} while the model is kept frozen. In DreamBooth \cite{ruiz2023dreambooth} the full weights of the model are finetuned while a prior preservation loss prevents the model from catastrophic forgetting and language drift \cite{lang_drift}. 
 CustomDiffusion \cite{kumari2022customdiffusion} and SVDiff \cite{han2023svdiff}
reduce the amount of fine-tuning parameters by only training the cross-attention layers.
While text-to-image personalization has been widely explored, the adoption of such methods for controllable music generation has not been addressed.

In this work, we start from a pre-trained text-to-audio diffusion model, i.e. \mbox{AudioLDM} \cite{liu2023audioldm}, and to the best of our knowledge are the first to investigate the ability to personalize its outputs for newly learned musical concepts in a few-shot manner.
Motivated by the computer vision literature, we explore the application of two established methods, i.e. Textual Inversion \cite{gal2023an} and Dreambooth \cite{ruiz2023dreambooth}.
We adapt these methods for music personalization and experiment with different training configurations. 
We evaluate the capacity of the model to learn new concepts along two dimensions, \emph{reconstruction}, i.e. the ability to faithfully reconstruct the novel concept, and \emph{editability}, i.e. the ability to manipulate it through textual prompts. 
To this end, we construct a new dataset of various instruments and playing styles. Our evaluation protocol consists of a) embedding distance-based metrics, b) music-specific metrics, and c) an A/B testing user study comparing the two adaptation approaches.
Finally, we adapt AudioLDM to perform text-guided style transfer for newly learned concepts. 

Our key contributions are a) the personalization of AudioLDM's generation and style-transfer abilities for new concepts, b) the exploration of audio-specific augmentations and evaluation metrics, and c) the construction of a new dataset for text-to-music personalization methods. Our code and data, as well as generated music samples, are publicly available \footnote{https://zelaki.github.io/}.

\section{Methods}
\label{sec:methods}

\noindent\textbf{Text-to-Audio Latent Diffusion Models:} 
Diffusion Models \cite{ho2020denoising} are probabilistic generative models
that learn a data distribution by gradually
denoising a latent variable sampled from a Gaussian distribution. 
This corresponds to learning the reverse process of a fixed-length
Markovian forward process. 

In Latent Diffusion Models (LDMs) \cite{ldm}, the denoising process occurs in the latent space of an encoder-decoder architecture ($\mathcal{E}$,$\mathcal{D}$) trained on a large collection of samples.
Given an audio sample $x$, a text-guided latent diffusion model is conditioned on a text-embedding model $c_{\tau}$. The LDM loss is then given by:
\begin{equation}
    \label{ldm} 
    \mathcal{L}_{LDM} = \mathbb{E}_{z \sim \mathcal{E}(x),\epsilon \sim \mathcal{N}(0, I), y, t} \left[  \| \epsilon - \hat{\epsilon}_{\phi}(z_t, t, c_{\tau}(y) \|_2^2 \right]
\end{equation}
where $\phi$,$\tau$ are the parameters of the denoising network $\hat{\epsilon}$ and the text encoder $c$ respectively, $t$ is the time step, $z_t$ is the latent representation of $x$ noised to time $t$ and $\epsilon$ is the sample noise.
While training, the parameters $\theta = \phi \; \cup \; \tau$  are jointly optimized to minimize the LDM loss.
Intuitively, the objective aims to correctly remove the noise added to a latent representation of
an audio. 
At inference, a
random noise tensor is sampled and iteratively denoised to produce a new audio latent, $z_0$, which is transformed into audio through the pre-trained decoder $\hat{x} = \mathcal{D}(z_0)$.

 \begin{figure}[ht!]
  \centering

  \includegraphics[scale=0.2]{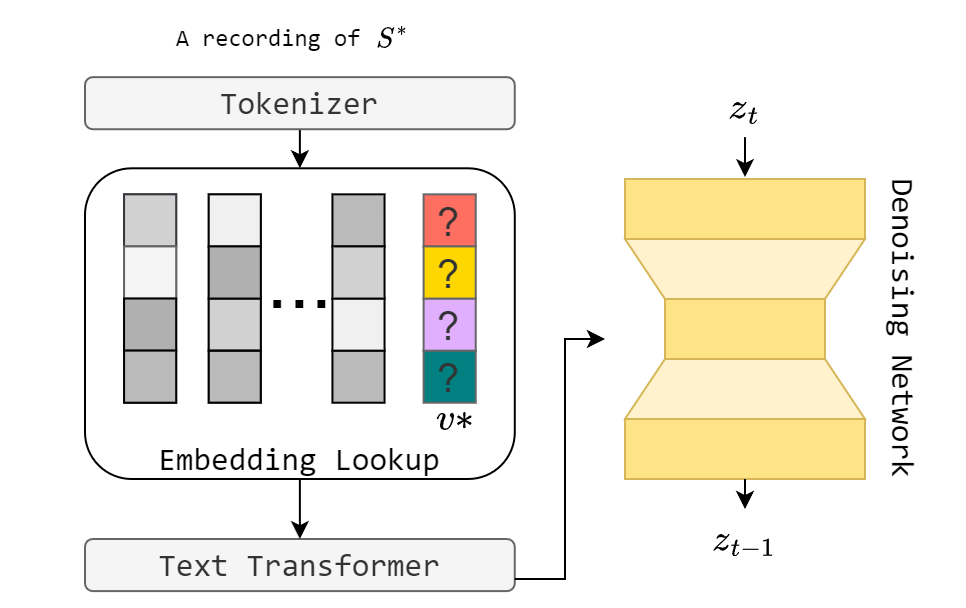}
  \caption{We illustrate the proposed methods for personalized text-to-music generation. In Textual Inversion only the novel embedding $v_*$ is optimized while in DreamBooth the full Denoising Network is finetuned.}
  
  \label{fig:model}
\end{figure}

\noindent\textbf{Personalization of Text-to-Audio Models:} 
Initially, a placeholder string, $S_*$, representing the novel concept is associated with a unique embedding vector  $v_*$  that can be retrieved via an embedding lookup as in Fig.\ref{fig:model}.  Thus the parameter space of the text encoder $c$ becomes $\tau'=\tau \; \cup \; v_*$ and the trainable parameters of the model become $\theta' = \phi \; \cup \; \tau \cup \; v_* $. The generation is conditioned to a constructed neutral text prompt $y$ e.g. "A recording of a $S_*$".

By directly minimizing the LDM loss (\ref{ldm}) over the small training set representing the concept and choosing different subsets of the parameter space of the model $\theta'$ for training, different methods for learning the novel concept can be derived. In Dreambooth (DB) the weights of the denoising network $\phi$ are optimized while in Textual Inversion (TI) $\phi$ and $\tau$ are kept frozen and the only learnable parameters are the weights of the embedding $v_*$.

\noindent\textbf{Personalized Style Transfer:} 
Given an input audio sample $x_{in}$, we can calculate its noisy latent representation $z_t$ with a predefined time step $t \leq N$ according to the forward process \cite{liu2023audioldm, ho2020denoising}.
By utilizing $z_t$
as the starting point
of the reverse process of a pre-trained AudioLDM model, we
enable the manipulation of $x_{in}$ with text input y
with a shallow reverse process:
\begin{equation}
    p_{\theta'}(z_{0:t} \vert c(y) = p(z_t)\prod_{n=1}^{t} p_{\theta'}(z_{n-1} \vert z_n, c(y)),
\end{equation}
where $t$ controls the transfer strength.
To infuse the style of the input sample $x_{in}$ with the characteristics of an acquired concept, we set $y = S_*$, where $S_*$ is the placeholder string linked to the newly learned concept.

\section{Experiments}
\label{sec:experiments}

\noindent\textbf{Dataset.} We collect a dataset of 32 musical concepts including Percussion Instruments and Beats, Solo Melodic Instruments, and Multi-Instrument Pieces, from a wide array of musical cultures and playing styles. Each concept includes five 10-second audio clips. All audio clips are either recorded by the authors or sourced from Freesound and YouTube. We also collect 20 editability prompts that aim to manipulate the genre, recording conditions, accompaniments, and background sounds. The full list of prompts can be found in the provided example page.

\noindent\textbf{Experimental Setup.} As a backbone for all our experiments we utilize AudioLDM-Medium \footnote{https://huggingface.co/cvssp/audioldm-m-full}.
We conduct our experiments using a single NVIDIA RTX-3090 GPU with a batch size of 4, employing learning rates of $2 \times 10^{-2}$ and $4 \times 10^{-6}$ for TI and DB, respectively, and running 150 optimization steps for TI and 1500 for DB.

Following the original papers for both methods, a placeholder token $S_*$ is reserved for the novel concept. In the case of TI, $S_*$ is a new token, inserted into the tokenizer, while for DB, \mbox{$S_*$ = [identifier] [class noun]}, where \mbox{[identifier]} is an existing rare word in the tokenizer and \mbox{[class noun]} is a coarse descriptor of the musical concept \cite{ruiz2023dreambooth}.
We experiment with different training configurations including training with 1 or 3 audio clips and randomly mixing the training audio with environmental sounds sourced from AudioSet \cite{audioset} with SNR$=20 \;$dB. We will refer to these experiments as 1-AC, 3-AC, and MIX respectively.

We further conduct ablation experiments specific to each method. 
For DB we include the text encoder in training, denoted as TE. For TI, we explore two possible initializations for the learnable embedding $v_*$.
As a baseline, we consider the initialization of $v*$ from the mean of all word embeddings in the vocabulary. Alternatively, we initialize $v*$ from the mean of the embeddings of \mbox{[class noun]}. We refer to the baseline and mean word initialization as BL and MW. For evaluation, we generate four 10-second music clips per concept and per prompt, totaling 2560 clips.


\noindent\textbf{Evaluation Metrics.} We evaluate the audio similarity between the training set and the music clips generated with a reconstruction prompt "a recording of a "$S_*$", by measuring \texttt{ CLAP-A} and \texttt{FAD} \cite{kilgour2019frechet} scores. \texttt{CLAP-A} is the average pairwise cosine similarity
between CLAP \cite{wu2023large} audio embeddings of generated and training clips.
 \texttt{FAD} is the Frechet distance between the embeddings of a pre-trained VGGish audio classifier of the training set and the generated music clips.
 We further evaluate the model's capacity to manipulate a learned audio concept via the editability prompts. For this, we calculate the average cosine similarity between the embedding of the editability prompt and the audio CLAP
embeddings and denote this metric as \texttt{CLAP-T}.
\begin{figure}[t]

  \centering

  \includegraphics[width=0.74\columnwidth]{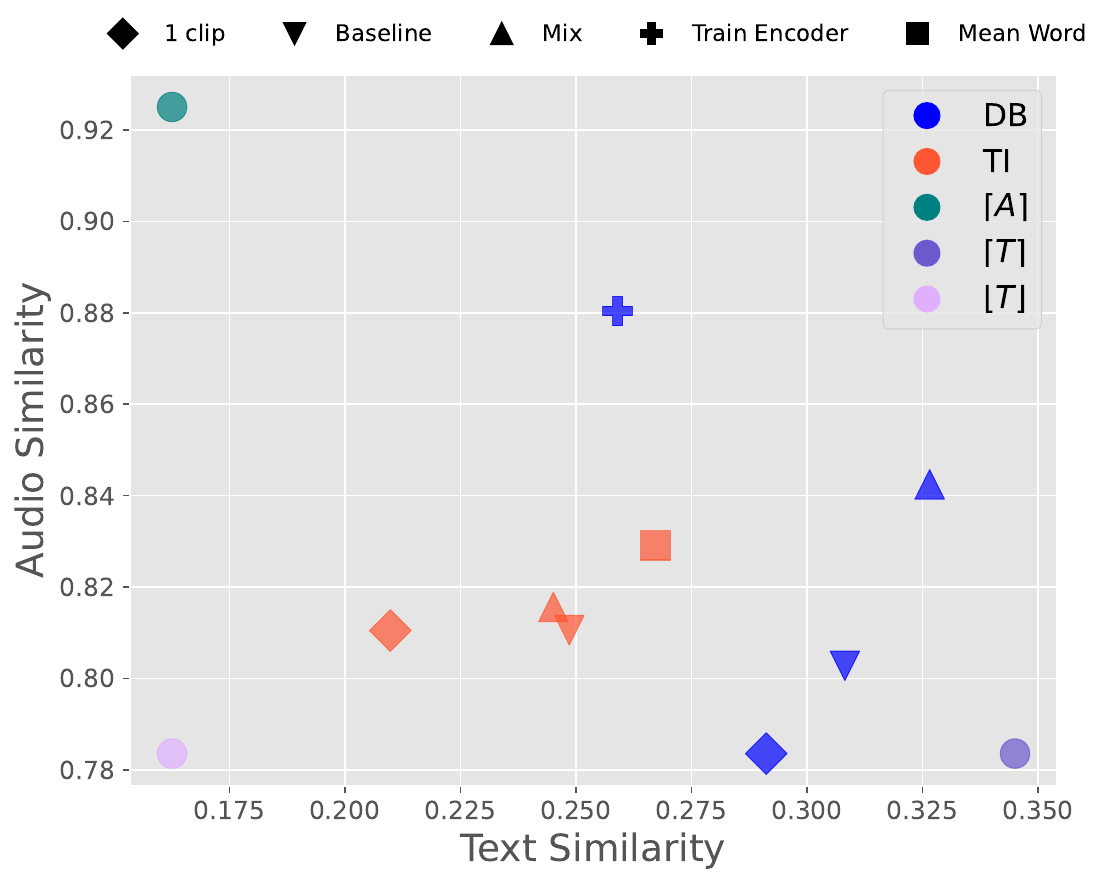}
  \caption{Audio and Text similarities for all experiments. TI methods (in orange) have roughly the same CLAP audio similarity but consistently lower text similarity than the DB ones.}  
  \label{fig:pareto_graph}
\vspace{-5 pt}
\end{figure}

Finally, we employ a set of automatically extracted music-specific metrics, in order to assess the capability of the model to retain certain musical properties, such as rhythm, harmony, and dynamics.
 To measure rhythmic similarity we compute the Beats per Minute (BPM) of the concepts that have a constant BPM and consider generated audio clips similar if they are within a 5 BPM tolerance \cite{Lee2019MusicSimilarity}. To measure similarity in dynamics, we use the European Broadcasting Union (EBU) R 128 Loudness Scale \cite{ebur128}, and consider clips similar in absolute loudness if they are within 2.5 LUFS of the mean training set loudness \cite{begnert2011difference}. Finally, to measure harmonic similarity, we use a key detection algorithm based on Harmonic Pitch Class Profiles (HPCP) \cite{demirel2019automatic}, which detects the fundamental tone of the clip's key, as well as whether it is major or minor. We compute the scale and key for all concepts that contain harmonic instruments and compare the generated clips' key and scale to the most common key and scale in the training set.
 All features are calculated using Essentia \cite{bogdanov2013essentia}.

\section{Results}

\noindent\textbf{Quantitative Analysis:} In Fig.~\ref{fig:pareto_graph}, we summarize our experimental results for the \texttt{CLAP-A} and \texttt{CLAP-T} metrics.
 To gain an intuition for the scale of the results, we add three references. The Audio Similarity ceiling, $\lceil A \rceil$ is the mean \texttt{CLAP-A} between the training samples of the concepts, the Text Similarity ceiling $\lceil T \rceil$ is the \texttt{CLAP-T} between the generated audio and the editability prompts without $S_*$ and the Text Similarity floor $\lfloor T \rfloor$ is the \texttt{CLAP-T} between the training audio and the editability prompts.
 We include  $\lceil T \rceil$ and $\lfloor T \rfloor$ scores to emphasize that the editability of the learned concepts is constrained by the prior manipulation capabilities of AudioLDM.
 
We observe that DB outperforms TI in terms of both similarity metrics.
Similar to the computer vision literature, we observe a ``Pareto front'' formed along the text and audio similarity axes, especially for DB \cite{gal2023encoder, tewel2023key}. 
When analyzing different training configurations, we observe that training using only one audio clip yields worse results both in terms of reconstruction (audio similarity) and in terms of editability (text similarity). 
Additionally, we observe that the MIX  strategy provides a good balance between reconstruction and editability. Further, MW outperforms the baseline TI, leveraging the prior of the embeddings of [class word]. Finally, training the text encoder overfits audio reconstruction while impeding the manipulation capacity through textual prompts.

In Table~\ref{tab:ablation}, we see a detailed view of the effect of different training configurations and include results on \texttt{FAD} score.
We observe that the audio reconstruction capability is strongly emphasized by the \texttt{FAD} score. While the MIX strategy has a significant impact on DB, it does not improve TI, since the regularization performed due to the data augmentation has a larger impact on the fully fine-tuned DB architecture.

\begin{table}[]
\centering

\label{tab:ablation}
\scalebox{0.75}{
\begin{tabular}{|c|c|c|c|c|}
\hline
\textbf{Method}              & \textbf{Setup} & \textbf{CLAP-A} ($\mathbf{\uparrow}$) & \textbf{FAD} ($\mathbf{\downarrow}$)  & \textbf{CLAP-T} ($\mathbf{\uparrow}$)\\ \hline
\multirow{5}{*}{DB} & BL    & $0.80$   & $12.37$ & $0.30$   \\
                    & 1-AC  & $0.78$   & $12.5$  & $0.29$   \\
                    & 3-AC  & $0.78$   & $13.04$ & $0.29$   \\
                    & TE    & $\mathbf{0.88}$   & $\mathbf{8.1}$   & $0.26$   \\
                    & MIX   & $0.84$   & $11.47$ & $\mathbf{0.33}$   \\ \hline
\multirow{5}{*}{TI} & BL    & $0.81$   & $17.17$ & $0.25$   \\
                    & 1-AC  & $0.810$  & $19.32$ & $0.21$   \\
                    & 3-AC  & $0.86$   & $17.45$ & $0.20$   \\
                    & MW    & $0.83$   & $16.61$ & $0.27$   \\
                    & MIX   & $0.816$  & $17.24$ & $0.25$   \\ \hline
\end{tabular}
}
\caption{Quantitative metrics comparing the different evaluation setups, for each method and training configuration.}
\vspace{-5 pt}
\end{table}

\begin{figure}[t]

  \centering
\hspace*{-0.5cm}
  \includegraphics[scale=0.285]{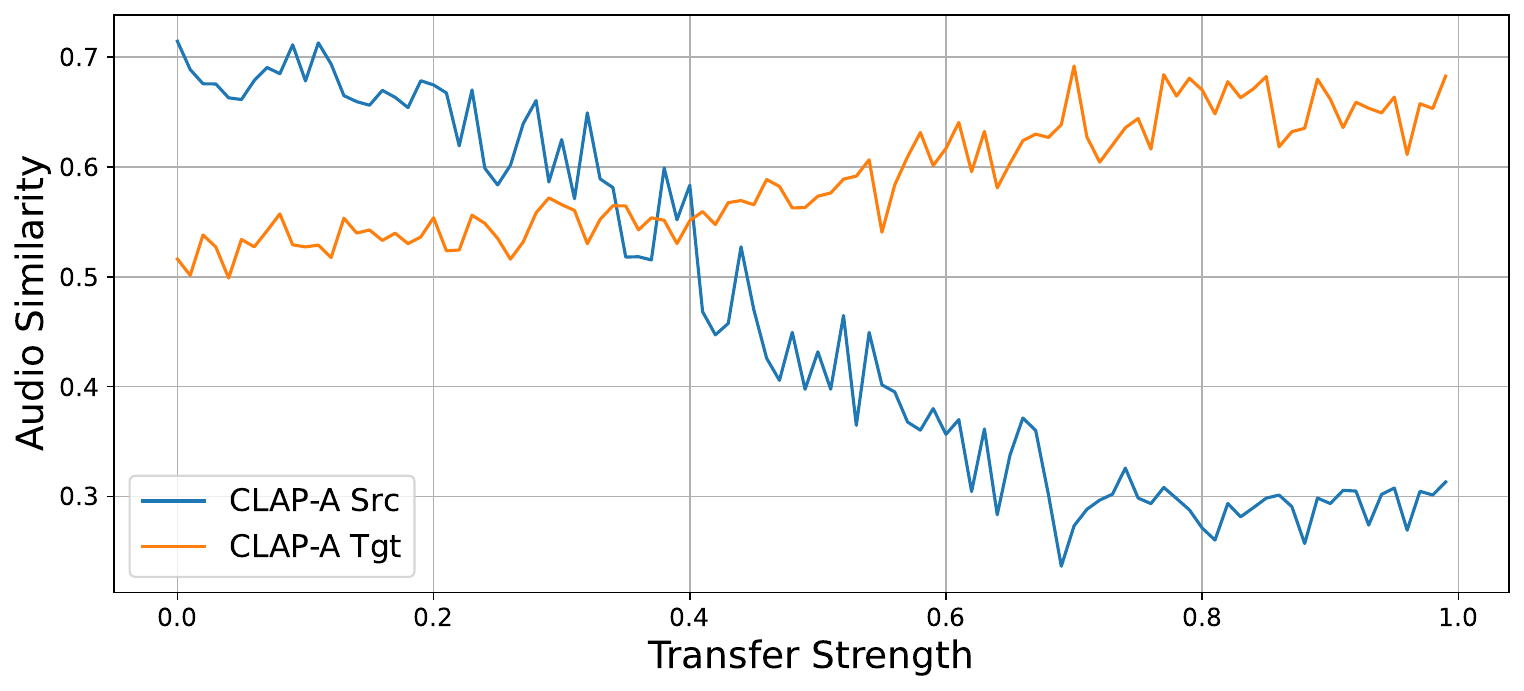}
  \caption{Audio Similarity for the input source and the target concept for personalized style transfer, with samples generated with transfer strength $t \in [ 0,1 ]$. The source is a 10-second clip from Stairway to Heaven and the target concept is a Middle Eastern string instrument called Oud.}  
  \label{fig:trasfer}
\vspace{-10 pt}
\end{figure}

\noindent\textbf{Human Preference Study:} We conduct a human preference study comparing DB and TI, in the form of an A/B testing setup, by creating an online survey consisting of $20$ questions. Half of the questions aim to evaluate the audio reconstruction ability, i.e. ``which of the two generated clips better matches the reference audio clip''. The other half, aim to evaluate the editability of DB and TI, by presenting the user with a textual prompt, including the novel concept class name, and then asking which of the generated clips better matches the prompt.
The possible preference answers for all questions are ``A'', ``B'', ``None'', and ``Cannot decide''. The study was completed by $34$ users. In Table~\ref{tab:userstudy} we can see that there is a preference towards DB instead of TI, in terms of reconstruction and editability in alignment with quantitative metrics. Furthermore, we observe that a significant amount of users do not prefer either DB or TI in the editability questions, indicating that text-to-music personalization still has room for improvement.

\begin{table}[]
\centering
\label{tab:userstudy}
\scalebox{0.9}{
\begin{tabular}{|c|c|c|c|c|}
\hline
               & \textbf{DB}   & \textbf{TI}   & \textbf{None} & \textbf{Undecided} \\ \hline
Reconstruction & $\mathbf{58}\textbf{\%}$ & $24$\% & $9$\%  & $9$\%      \\ \hline
Editability    & $\mathbf{37}$\textbf{\%} & $31$\% & $24$\% & $8$\%       \\ \hline
\end{tabular}
}
\caption{Human preference study results.}

\vspace{-10 pt}
\end{table}



\begin{figure}[h]
\vspace{-4 pt}

  \centering

  \includegraphics[scale=0.28]{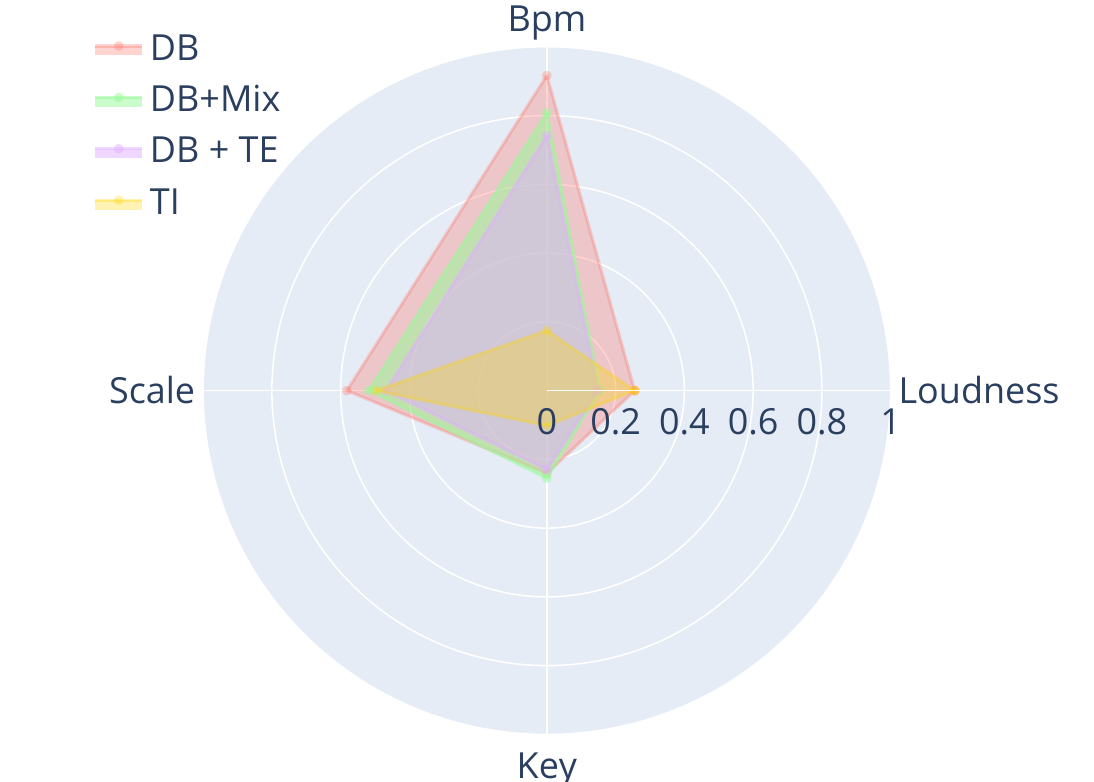}
  \caption{Low-level music audio features for some of the training configurations.}
  \label{fig:music_metrics}
\vspace{-5 pt}
\end{figure}

\noindent\textbf{Music-Specific Evaluation:} In Fig.~\ref{fig:music_metrics} we illustrate the results for DB in three training configurations and the baseline TI on the proposed music metrics. Initially, we observe that DB can effectively retain tempo, while TI cannot. Additionally, while both methods are able to reconstruct the scale to some extent they fail to reconstruct the key.
Finally, both methods cannot generate clips with loudness comparable to the training set. We hypothesize that this is due to the model's normalizing effect on the generated audio.

\noindent\textbf{Text-to-Audio Style-Transfer:} In Fig.~\ref{fig:trasfer} we perform personalized style-transfer for a novel concept, using TI+MW. 
We present that by increasing transfer strength the generated audio progressively becomes similar to the target clip, and dissimilar to the source clip. Furthermore, the range $0.4-0.6$ for the transfer strength appears to be a sweet spot for performing style-transfer, while maintaining the source audio properties.


\section{Conclusion and Future Work}
\label{sec:future_work}

In this work, we conduct a preliminary study for the personalization of text-to-music generation models adapting them
to user-specific needs. We explore the application of two established methods, namely Textual Inversion and DreamBooth. Both methods are evaluated on
their ability to learn and modify new musical concepts, using
quantitative metrics and a user study. We construct a new evaluation dataset, investigate diverse training configurations, and propose a music-specific evaluation framework.
In the future, we aim to explore multi-concept text-to-music personalization, learning multiple music concepts from a single mixture, with a source separation regularization objective. 


\vfill\pagebreak

\bibliographystyle{IEEEbib}
\bibliography{Template}

\begin{thebibliography}{10}

\bibitem{liu2023audioldm}
Haohe Liu, Zehua Chen, Yi~Yuan, Xinhao Mei, Xubo Liu, Danilo Mandic, Wenwu
  Wang, and Mark~D Plumbley,
\newblock ``Audioldm: Text-to-audio generation with latent diffusion models,''
\newblock {\em arXiv preprint arXiv:2301.12503}, 2023.

\bibitem{copet2023simple}
Jade Copet, Felix Kreuk, Itai Gat, Tal Remez, David Kant, Gabriel Synnaeve,
  Yossi Adi, and Alexandre D{\'e}fossez,
\newblock ``Simple and controllable music generation,''
\newblock {\em arXiv preprint arXiv:2306.05284}, 2023.

\bibitem{gu2023systematic}
Jindong Gu, Zhen Han, Shuo Chen, Ahmad Beirami, Bailan He, Gengyuan Zhang,
  Ruotong Liao, Yao Qin, Volker Tresp, and Philip Torr,
\newblock ``A systematic survey of prompt engineering on vision-language
  foundation models,'' 2023.

\bibitem{ruiz2023dreambooth}
Nataniel Ruiz, Yuanzhen Li, Varun Jampani, Yael Pritch, Michael Rubinstein, and
  Kfir Aberman,
\newblock ``Dreambooth: Fine tuning text-to-image diffusion models for
  subject-driven generation,''
\newblock in {\em In Proc. CVPR}, 2023.

\bibitem{gal2023an}
Rinon Gal, Yuval Alaluf, Yuval Atzmon, Or~Patashnik, Amit~Haim Bermano, Gal
  Chechik, and Daniel Cohen-or,
\newblock ``An image is worth one word: Personalizing text-to-image generation
  using textual inversion,''
\newblock in {\em The Eleventh International Conference on Learning
  Representations}, 2023.

\bibitem{kumari2022customdiffusion}
Nupur Kumari, Bingliang Zhang, Richard Zhang, Eli Shechtman, and Jun-Yan Zhu,
\newblock ``Multi-concept customization of text-to-image diffusion,''
\newblock 2023.

\bibitem{ldm}
Robin Rombach, Andreas Blattmann, Dominik Lorenz, Patrick Esser, and Bj{\"o}rn
  Ommer,
\newblock ``High-resolution image synthesis with latent diffusion models,''
\newblock in {\em In Proc. CVPR}, 2022, pp. 10684--10695.

\bibitem{lang_drift}
Jason Lee, Kyunghyun Cho, and Douwe Kiela,
\newblock ``Countering language drift via visual grounding,''
\newblock {\em arXiv preprint arXiv:1909.04499}, 2019.

\bibitem{han2023svdiff}
Ligong Han, Yinxiao Li, Han Zhang, Peyman Milanfar, Dimitris Metaxas, and Feng
  Yang,
\newblock ``Svdiff: Compact parameter space for diffusion fine-tuning,'' 2023.

\bibitem{ho2020denoising}
Jonathan Ho, Ajay Jain, and Pieter Abbeel,
\newblock ``Denoising diffusion probabilistic models,''
\newblock {\em Advances in neural information processing systems}, vol. 33, pp.
  6840--6851, 2020.

\bibitem{audioset}
Jort~F. Gemmeke, Daniel P.~W. Ellis, Dylan Freedman, Aren Jansen, Wade
  Lawrence, R.~Channing Moore, Manoj Plakal, and Marvin Ritter,
\newblock ``Audio set: An ontology and human-labeled dataset for audio
  events,''
\newblock in {\em In Proc ICASSP}, 2017, pp. 776--780.

\bibitem{kilgour2019frechet}
Kevin Kilgour, Mauricio Zuluaga, Dominik Roblek, and Matthew Sharifi,
\newblock ``Fr{\'e}chet audio distance: A reference-free metric for evaluating
  music enhancement algorithms.,''
\newblock 2019.

\bibitem{wu2023large}
Yusong Wu, Ke~Chen, Tianyu Zhang, Yuchen Hui, Taylor Berg-Kirkpatrick, and
  Shlomo Dubnov,
\newblock ``Large-scale contrastive language-audio pretraining with feature
  fusion and keyword-to-caption augmentation,''
\newblock in {\em In Proc. ICASSP}. IEEE, 2023, pp. 1--5.

\bibitem{Lee2019MusicSimilarity}
Jongpil Lee, Nicholas~J. Bryan, Justin Salamon, Zeyu Jin, and Juhan Nam,
\newblock ``Disentangled multidimensional metric learning for music
  similarity,''
\newblock in {\em In Proc. ICASSP}. IEEE, 2020.

\bibitem{ebur128}
European~Broadcasting Union,
\newblock ``Loudness normalisation and permitted maximum level of audio
  signals,'' 2020.

\bibitem{begnert2011difference}
Fabian Begnert, H{\aa}kan Ekman, and Jan Berg,
\newblock ``Difference between the ebu r-128 meter recommendation and human
  subjective loudness perception,''
\newblock in {\em Audio Engineering Society Convention 131}. Audio Engineering
  Society, 2011.

\bibitem{demirel2019automatic}
Emir Demirel, Baris Bozkurt, and Xavier Serra,
\newblock ``Automatic chord-scale recognition using harmonic pitch class
  profiles,''
\newblock in {\em Barbancho I, Tard{\'o}n LJ, Peinado A, Barbancho AM, editors.
  Proceedings of the 16th Sound \& Music Computing Conference; 2019 May 28-31;
  M{\'a}laga, Spain.[M{\'a}laga]: SMC; 2019.} Sound \& Music Computing
  Conference, 2019.

\bibitem{bogdanov2013essentia}
Dmitry Bogdanov, Nicolas Wack, Emilia G\'{o}mez, Sankalp Gulati, Perfecto
  Herrera, Oscar Mayor, Gerard Roma, Justin Salamon, Jos\'{e} Zapata, and
  Xavier Serra,
\newblock ``Essentia: An open-source library for sound and music analysis,''
\newblock in {\em Proceedings of the 21st ACM International Conference on
  Multimedia}, New York, NY, USA, 2013, MM '13, p. 855–858, Association for
  Computing Machinery.

\bibitem{gal2023encoder}
Rinon Gal, Moab Arar, Yuval Atzmon, Amit~H Bermano, Gal Chechik, and Daniel
  Cohen-Or,
\newblock ``Encoder-based domain tuning for fast personalization of
  text-to-image models,''
\newblock {\em ACM Transactions on Graphics (TOG)}, vol. 42, no. 4, pp. 1--13,
  2023.

\bibitem{tewel2023key}
Yoad Tewel, Rinon Gal, Gal Chechik, and Yuval Atzmon,
\newblock ``Key-locked rank one editing for text-to-image personalization,''
\newblock in {\em ACM SIGGRAPH 2023 Conference Proceedings}, 2023, pp. 1--11.

\end{thebibliography}

\end{document}